\documentclass{article} 
\usepackage{nips13submit_e,times}
\usepackage{hyperref}
\usepackage{url}

\usepackage{epsfig}
\usepackage{amsmath}
\usepackage{amsthm}
\usepackage{amssymb}
\usepackage{algorithm}
\usepackage{algorithmic}

\usepackage[page]{appendix}

\newcommand{\eq}{\begin{equation*}}
\newcommand{\en}{\end{equation*}}
\newcommand{\eqa}{\begin{eqnarray*}}
\newcommand{\ena}{\end{eqnarray*}}
\newcommand{\eqn}{\begin{equation}}
\newcommand{\enn}{\end{equation}}
\newcommand{\be}{\begin{equation}}
\newcommand{\ee}{\end{equation}}
\newcommand{\eqan}{\begin{eqnarray}}
\newcommand{\enan}{\end{eqnarray}}

\newcommand{\vp}{\varpropto}
\newcommand{\nn}{\nonumber}

\newcommand{\y}{ {\bf y} }

\newcommand{\st}{ {\bf s} }
\newcommand{\sm}{\st^{{}_{-}}}
\newcommand{\spl}{\st^{{}_{+}}}

\newcommand{\q}{ {\bf q} }
\newcommand{\g}{ {\bf g} }
\newcommand{\w}{ {\bf w} }
\newcommand{\x}{ {\bf x} }

\newcommand{\z}{ {\bf z} }

\newcommand{\pmat}{\begin{pmatrix}}
\newcommand{\pman}{\end{pmatrix}}

\title{ Auxiliary-variable Exact Hamiltonian Monte \\ Carlo Samplers for Binary Distributions} 

\author{
Ari Pakman and Liam Paninski 
\\
Department of Statistics
\\
Center for Theoretical Neuroscience
\\
Grossman Center for the Statistics of Mind
\\
Columbia University
\\
New York, NY, 10027
}

\nipsfinalcopy 

\begin{document}

\maketitle

\begin{abstract}
We present a new approach to sample from generic binary distributions, based on 
an exact  Hamiltonian Monte Carlo algorithm applied
to a piecewise continuous augmentation  of the binary distribution of interest. 
An extension of this idea to distributions over mixtures of binary and possibly-truncated  Gaussian or exponential variables
allows us to sample from posteriors of  linear and probit regression models with  spike-and-slab priors and truncated parameters.  
We illustrate the advantages of these algorithms in several examples in which 
they outperform the Metropolis or Gibbs samplers. 
\end{abstract}

\section{Introduction}
Mapping a  problem involving discrete variables into continuous variables often results in a more tractable formulation. 
For the case of probabilistic inference, in this paper we present a new approach to sample from distributions over binary variables, based on mapping
the original discrete distribution into a continuous one with a piecewise quadratic log-likelihood, from which we can sample efficiently using 
exact Hamiltonian Monte Carlo (HMC). 

The HMC method is a Markov Chain Monte Carlo algorithm that usually has  better performance over Metropolis or Gibbs samplers, 
because it manages to propose transitions  in the Markov chain which lie far apart in the sampling space, while maintaining a
reasonable acceptance rate for these proposals.    But the implementations of HMC algorithms  generally involve the non-trivial tuning 
of numerical integration parameters  to obtain such a reasonable acceptance rate (see~\cite{neal2011mcmc} for a review).
The algorithms we present in this work are special because the Hamiltonian equations of motion can be integrated exactly, so there is no 
 need for tuning a step-size parameter and the Markov chain always accepts the  proposed moves. Similar ideas have been used recently to sample
from truncated Gaussian multivariate distributions~\cite{pakman2013exact}, allowing much faster sampling than other methods.
It should be emphasized that despite the apparent complexity of deriving the new algorithms, 
their implementation is very simple.

Since the method we present transforms a binary sampling problem into a continuous one, it is natural to extend it to distributions
defined over mixtures of binary and Gaussian or exponential variables, transforming them into purely continuous distributions. 
Such a mixed binary-continuous problem arises in Bayesian model selection with a  spike-and-slab prior
and we illustrate our technique by focusing on this case. In particular, we show how to  sample from the posterior of linear and probit regression models 
with  spike-and-slab priors, while  also imposing truncations in the parameter space (e.g., positivity).

The method we use to map binary to continuous variables consists in simply identifying a binary variable with the sign of a continuous one.
An alternative relaxation of binary to continuous variables, known in statistical physics as the ``Gaussian integral trick''~\cite{hertz1991introduction},
 has been used recently to apply HMC methods to binary distributions~\cite{zhang2012continuous}, but the details of 
that method are different than ours. In particular, the HMC in that work  is not `exact' in the sense used above and the algorithm only works for 
Markov random fields with Gaussian potentials.

\section{Binary distributions}
We are interested in sampling from a  probability  distribution $p(\st)$ defined over $d$-dimensional binary vectors $\st \in \{-1,+1\}^{d}$,
and given in terms of a function $f(\st)$ as
\eqan 
p(\st) = \frac{1}{Z}f(\st) \,.
\enan 
Here $Z$ is a normalization factor, whose value will not be needed. 
Let us augment the distribution~$p(\st)$ with continuous variables $\y \in \mathbb{R}^{d}$ as
\eqan
p(\st,\y)  = p(\st)p( \y |\st)
\label{psy}
\enan 
where $p( \y |\st)$ is non-zero only in the orthant defined by 
\eqan
s_i = sign(y_i) \qquad \qquad  i =1, \ldots, d.
\label{sy}
\enan
The essence of the proposed method is that we can sample from $p(\st)$  by sampling $\y$ from 
\eqan
p(\y)  &=& \sum_{\st'} p(\st') p(\y|\st') \,,
\label{ysum}
\\
&=& p(\st) p(\y|\st) \,,
\label{ynosum}
\enan 
and reading out the values of $\st$ from  (\ref{sy}). 
In the second line  we have made explicit that  for each  $\y,$ only one term in the sum in~(\ref{ysum}) is non-zero, 
so that $p(\y)$ is  piecewise defined in each orthant.

In order to sample from $p(\y)$ using the exact HMC method of~\cite{pakman2013exact}, 
we require $\log p(\y|\st)$ to be  a quadratic function of $\y$ on its support.
The idea is to define a potential energy function
\eqan
U(\y) = - \log p(\y|\st)  - \log f(\st) \,,
\enan 
introduce momentum variables $q_i$, and consider the piecewise continuous Hamiltonian
\eqan 
H(\y, \q) &= U(\y) + \frac{\q \cdot \q}{2} \,,
\label{Hh}
\enan 
whose value is identified with the energy of a particle moving in a $d$-dimensional space.
Suppose the  particle has initial coordinates $\y(0)$.  In each iteration of the sampling algorithm, we sample initial values $\q(0)$ for the momenta from a standard Gaussian distribution
and let the particle move during a time $T$ according to the equations of motion 
\eqan
\dot{\y}(t) = \frac{\partial H}{\partial \q(t)} \,, \qquad \quad \dot{\q}(t) = - \frac{\partial H}{\partial \y (t)} \,.
\label{eom}
\enan 
The final coordinates, $\y(T)$, belong to a Markov chain with invariant distribution $p(\y)$, and are used as the initial coordinates of the next iteration.
The detailed balance condition follows directly from the conservation of energy and  $(\y,\q)$-volume along the trajectory dictated by (\ref{eom}), 
see~\cite{neal2011mcmc,pakman2013exact} for details.

The restriction to quadratic functions of $\y$ in $\log p(\y|\st)$ allows us to solve the differential equations~(\ref{eom}) exactly in each orthant.
As the particle moves, the potential energy $U(\y)$ and the kinetic energy $\frac{\q \cdot \q}{2}$ change in tandem,  keeping
the value of the Hamiltonian (\ref{Hh}) constant. But this smooth interchange gets interrupted when  any coordinate reaches zero.
Suppose this first happens at time~$t_j$ for coordinate $y_j$, and assume that 
$y_j < 0$ for $t < t_j$.  
Conservation of energy imposes now a jump on the momentum $q_j$ as a result of the  discontinuity in $U(\y)$.
Let us call $q_j(t_j^-)$ and~$q_j(t_j^+)$ the value of the momentum $q_j$ just before and after the coordinate hits $y_j=0$.
In order to enforce conservation of energy, we equate the Hamiltonian at both sides of the $y_j=0$ wall,  giving
\eqan 
\frac{q_j^2(t_j^+)}{2} = \Delta_j + \frac{q_j^2(t_j^-)}{2}
\label{kin}
\enan 
with 
\eqan 
\Delta_j = U( y_j=0,  s_{j}=-1) - U( y_j=0,  s_{j}=+1)
\label{deltaj}
\enan 
If eq. (\ref{kin}) gives a positive value for $q_j^2(t_j^+)$, the coordinate $y_j$ crosses the boundary and continues its trajectory in the new orthant.
On the other hand, if eq.(\ref{kin})  gives a negative value for~$q_j^2(t_j^+)$, the particle is reflected from the $y_j=0$ wall 
and continues its trajectory with $q_j(t_j^+) = - q_j(t_j^-)$.
When $\Delta_j < 0$, the situation  can be understood as the limit of a scenario in which the particle faces an upward hill in the potential energy,  causing it to diminish its velocity 
until it either reaches the top of the hill with a lower velocity  or stops and then reverses. In the limit in which the hill has finite height but infinite slope, 
the velocity change  occurs discontinuously at one instant.  Note that  we used in (\ref{kin}) that the momenta $q_{i\neq j}$ are continuous, since this sudden 
infinite slope hill is only seen by the $y_j$ coordinate.

Regardless of whether the particle bounces or crosses the $y_j=0$ wall,  the other coordinates move  unperturbed until the next boundary hit, 
where a similar crossing or reflection occurs, and so on, until the final position $\y(T)$.

The framework we presented above is very general and in order to implement a  particular sampler we need to 
select the distributions~$p(\y|\st)$. Below we consider in some detail two particularly simple choices that illustrate the diversity of options here.

\subsection{Gaussian augmentation}
Let us consider first for $p(\y|\st)$ the truncated Gaussians
\eqan 
p(\y|\st) = 
\left\{
 \begin{array}{ll}
(2/\pi)^{d/2}\, e^{- \frac{\y \cdot \y}{2}}   &  \textrm{for} \,\,  sign(y_i)=s_i, \qquad  i =1, \ldots, d
\\
0 & \textrm{otherwise} \,,
\end{array}
\right.  
\label{pys}
\enan 
The equations of motion (\ref{eom}) lead to $\ddot{\y}(t) = - \y(t), \ddot{\q}(t) = - \q(t)$, and have a solution
\eqan
y_i(t) &=& y_i(0)\cos(t) + q_i(0) \sin(t) \,,
\label{yt}
\\
&=& u_i \sin(\phi_i + t) \,,
\label{yt2}
\\
q_i(t) &=& -y_i(0)\sin(t) + q_i(0) \cos(t) \,,
\\
&=& u_i \cos(\phi_i + t) \,.
\enan 
This setting is similar to the case studied in~\cite{pakman2013exact} and from  $\phi_i = \tan^{-1}( y_i(0)/q_i(0) )$ the boundary hit times $t_i$ are easily obtained.
When a boundary is reached, say $y_j=0$,  the coordinate $y_j$ changes its trajectory for $t > t_j$ as 
\eqan
y_j(t) &=& q_j(t_j^+) \sin(t-t_j)  \,,
\label{yj}
\enan 
with the value of $q_j(t_j^+)$ obtained as described above.

Choosing an appropriate value for the travel time $T$ is crucial when using HMC  algorithms~\cite{hoffman2011no}.
As is clear from (\ref{yt2}), if we let the particle travel during a time $T>\pi$, each coordinate reaches zero at least once, and the hitting times can be ordered as 
\eqan
0 < t_{j_1} \leq t_{j_2} \leq \cdots \leq t_{j_d} < \pi \,.
\label{cyc}
\enan 
Moreover, regardless of whether a coordinate crosses zero or gets reflected, it follows from (\ref{yj}) that the successive hits  occur at 
\eqan
t_{i} + n\pi, \quad n=1,2,\ldots
\enan 
and therefore the hitting times only need to be computed once for each coordinate in every iteration.
If we let the particle move during a time $T=n\pi$, each coordinate reaches zero $n$ times, in the cyclical order (\ref{cyc}),  
with  a  computational cost of $O(nd)$ from wall hits.  But choosing \emph{precisely} $T=n\pi$ is not recommended 
for the following reason. As we just showed, between $y_j(0)$ and $y_j(\pi)$ the coordinate touches the boundary $y_j=0$ once,
and if $y_j$ gets reflected off the boundary, it is easy to check that we have $y_j(\pi) = y_j(0)$. 
If we take $T=n\pi$ and the particle gets reflected all the $n$ times it hits the boundary, 
we get $y_j(T) = y_j(0)$ and the coordinate $y_j$  does not move at all. To avoid these singular situations, a good choice is~$T=(n + \frac12) \pi$,
which generalizes the recommended travel time $T=\pi/2$ for truncated Gaussians in~\cite{pakman2013exact}.
The value of $n$ should be chosen for each distribution, but we expect  optimal values for $n$ to grow with $d$.

With $T=(n + \frac12) \pi$, the total cost of each sample is $O((n+1/2)d)$ on average from wall hits,
plus $O(d)$ from  the sampling of $\q(0)$ and from the $d$ inverse trigonometric functions to obtain the hit times $t_i$.
But in complex distributions, the computational cost is dominated by the the evaluation of~$\Delta_i$ in (\ref{deltaj}) at each wall hit. 

Interestingly, the rate at which wall $y_i=0$ is crossed coincides with the acceptance rate in a Metropolis algorithm that 
samples uniformly a value for $i$ and makes a proposal of flipping the binary variable~$s_i$. 
See the Appendix for details. 
Of course, this does not mean that the HMC algorithm is the same as Metropolis, because  in HMC the order in which the walls are hit is fixed given the initial velocity,
and the values of $q^2_i$ at successive hits of $y_i=0$ within the same iteration  are not independent.

\subsection{Exponential and other augmentations}
Another distribution that allows one an exact solution of the equations of motion (\ref{eom}) is
\eqan 
p(\y|\st) = \left\{
 \begin{array}{ll}
 e^{- \sum_{i=1}^d |y_i|}   &  \textrm{for} \,\,  sign(y_i)=s_i, \qquad  i =1, \ldots, d
\\
0 & \textrm{otherwise} \,,
\end{array}
\right.  
\label{pyse}
\enan 
which leads to the equations of motion $\ddot{y_i}(t) = -s_i$, with solutions of the form 
\eqan 
y_i(t) &=& y_i(0) + q_i(0)t -\frac{s_it^2}{2} \,.
\enan 
In this case, the initial hit time for every coordinate is the solution of the quadratic equation~$y_i(t)=0$.
But, unlike the case of the Gaussian augmentation, the order of successive hits is not fixed. Indeed, if coordinate $y_j$ hits zero at time $t_j$, 
it continues its trajectory as
\eqan 
y_j(t> t_j) = q(t_j^+)(t-t_j) - \frac{s_j}{2}(t-t_j)^2 \,,
\enan 
so the next wall hit $y_j=0$ will occur  at a time $t'_j$ given by
\eqan
(t'_j-t_j) = 2|q_j(t_j^+)| \,,
\label{nextt}
\enan 
where we used $s_j = sign(q_j(t_j^+))$. So we see that the time between successive hits of the same coordinate depends only on its momentum 
 after the last hit. Moreover, since the value of $|q_j(t^+)|$ is smaller than $|q_j(t^-)|$ if the coordinate crosses to an orthant of lower probability, 
 equation (\ref{nextt}) implies that the particle moves away faster from areas of lower probability.
 This is unlike the Gaussian augmentation, where a coordinate `waits in line' until all the other coordinates 
 touch their wall before hitting its wall again.

The two augmentations we considered above have only scratched the surface of interesting possibilities.
One could also  define $f(\y|\st)$ as a uniform distribution in a box such that the computation of the times for wall hits 
would becomes purely linear and we get a classical `billiards' dynamics.
 More generally, one could consider different augmentations in different orthants 
and  potentially tailor the algorithm to mix faster in complex and multimodal  distributions.

%
%

\section{Spike-and-slab regression with truncated parameters}  

The subject of Bayesian sparse regression has seen a lot of work during the last decade.
Along with priors such as the Bayesian Lasso~\cite{park2008bayesian} and the Horsehoe~\cite{carvalho2010horseshoe}, 
the classic spike-and-slab prior~\cite{mitchell1988bayesian, george1993variable}
still remains very competitive, as shown by its superior performance in the recent works~\cite{mohamed2011bayesian,goodfellow2012spike, chen2012bayesian}.
But despite its successes, posterior inference  remains a computational challenge for the spike-and-slab prior.
In this section we will show how the HMC binary sampler can be extended to sample from the posterior of these models.
The latter is a distribution over a set of binary and continuous variables, with the 
binary variables determining whether each coefficient should be included in the model or not. 
The idea is to map these indicator binary variables into continuous variables as we did above, obtaining a 
distribution from which we can sample again using exact HMC methods.
Below we consider a regression model with Gaussian noise but the extension to exponential noise (or other scale-mixtures of Gaussians) is immediate.

\subsection{Linear regression}
Consider a regression problem with a log-likelihood that depends quadratically on its coefficients, such as  
\eqan 
\log p(D|\w) = -\frac12 \w'{\bf M} \w + {\bf r}\cdot \w + const.
\enan
where $D$ represents the observed data. In a linear regression model ${\bf z} = X\w + {\bf \varepsilon}$, with ${\bf \varepsilon} \sim {\cal N}(0,\sigma^2)$,  
we have ${\bf M} = X'X/\sigma^2$ and ${\bf r}={\bf z}'X/\sigma^2$. 
We are interested in a spike-and-slab prior for the coefficients $\w$ of the form
\eqan
p(\w,\st|a,\tau^2) = \prod_{i=1}^{d} p(w_i|s_i,\tau^2) p(s_i|a) \,.
\label{sns}
\enan 
Each binary variable $s_i = \pm 1$ has a Bernoulli prior $p(s_i|a) = a^{\frac{(1+s_i)}{2}} (1-a)^{\frac{(1-s_i)}{2}} $
and determines whether the coefficient $w_i$ is included in the model. The prior  for $w_i$, conditioned on $s_i$, is 
\eqan
p(w_i|s_i,\tau^2) = \left\{
 \begin{array}{ll}
\frac{1}{\sqrt{2 \pi \tau^2}}  e^{- \frac{w_i^2}{2 \tau^2}}  &  \textrm{for} \,\,  s_i=+1,
\label{slab}
\\
\\
\delta(w_i) & \textrm{for} \,\,  s_i=-1
\end{array}
\right.  
\enan 
We are interested in sampling from the posterior, given by 
\eqan 
p(\w,\st|D,a,\tau^2)  &\vp & p(D|\w) p(\w,\st|a,\tau^2) 
\\
&\vp &  
\frac{ 
e^{-\frac12  \w' {\bf M}  \w   + {\bf r} \cdot \w } 
e^{-\frac12  \w'_{+}\w_{+}  \tau^{-2} }}
          {(2 \pi \tau^2)^{|\spl|/2}}  
          \delta(\w_{-}) a^{|\spl|} (1-a)^{|\sm|}
\label{ssz}
\\
&\vp& 
\frac{ 
e^{-\frac12  \w'_{+} \left( {\bf M}_{+} + \tau^{-2}  \right) \w_{+}   + {\bf r}_{+}\cdot \w_{+} }}
          {(2 \pi \tau^2)^{|\spl|/2}}  
          \delta(\w_{-}) a^{|\spl|} (1-a)^{|\sm|}
\label{sso}
\enan
where $\spl$ are the variables with $s_i=+1$ and $\sm$  those with $s_i=-1$. The notation~${\bf r}_{+}$,~${\bf M}_{+}$ and $\w_{+}$  indicates a restriction
to the $\spl$ subspace and $\w_{-}$ indicates a restriction to the $\sm$ space. 
In the passage from (\ref{ssz}) to (\ref{sso}) we see that the spike-and-slab prior shrinks the dimension of the Gaussian
likelihood from $d$ to $|\spl|$. 
In principle we could integrate out the weights $\w$ and 
obtain a collapsed distribution for $\st$, but we are interested in cases in which the space of $\w$ is truncated and therefore the integration is not feasible.
An example would be when a non-negativity  constraint $w_i \geq 0$ is imposed. 

In these cases, one possible approach is to sample from~(\ref{sso}) with a block Gibbs sampler over the pairs~$\{w_i,s_i\}$, as proposed in~\cite{mohamed2011bayesian}. 
Here we will present an alternative method, extending the ideas of the previous section. For this, we consider a new distribution, obtained in two steps.
Firstly, we replace the delta functions  in (\ref{sso}) by a factor similar to the slab (\ref{slab})
\eqan 
\delta(w_i) \rightarrow  \frac{1}{\sqrt{2 \pi \tau^2}}e^{- \frac{w_i^2}{2 \tau^2}} \qquad \qquad s_i = -1
\label{rjs}
\enan 
The  introduction of a non-singular distribution for those  $w_i$'s that are excluded from the model in~(\ref{sso}) 
creates a Reversible Jump sampler~\cite{green1995reversible}: 
the Markov chain can now keep track of all the coefficients, whether they belong or not to the model in a given state of the chain,
thus allowing them to join or leave the model along the chain in a \emph{reversible} way.

Secondly,  we augment the distribution with $\y$ variables as in (\ref{psy})-(\ref{ynosum}) and sum over $\st$. 
Using the Gaussian augmentation (\ref{pys}), this gives a distribution
\eqan 
p(\w,\y|D,a,\tau^2)  \vp  e^{-\frac12  \w'_{+} \left( {\bf M}_{+} + \tau^{-2} \right) \w_{+}   + {\bf r}_{+}\cdot \w_{+} }            
           e^{-\frac{\w_{-}\cdot \w_{-} }{2 \tau^2}} e^{-\frac{\y \cdot \y}{2}}  a^{|\spl|} (1-a)^{|\sm|}
\label{ssn}
\enan 
where the values of $\st$ in the rhs are obtained from the signs of $\y$.
This is a piecewise  Gaussian, different in each orthant of $\y$, and possibly truncated in the $\w$ space. 
Note that the changes in $p(\w,\y|D,a,\tau^2)$ across orthants of $\y$ come both  from  the factors $a^{|\spl|} (1-a)^{|\sm|}$ and  from the functional dependence on the $\w$  variables.
Sampling from~(\ref{ssn}) gives us samples from the original distribution (\ref{sso}) using a simple rule: each pair $(w_i,y_i)$ becomes~$(w_i,s_i =+1)$ if $y_i \geq 0$ and 
$(w_i=0,s_i =-1)$ if $y_i < 0$. 
This undoes the steps we took to transform (\ref{sso}) into (\ref{ssn}): the identification $s_i = sign(y_i)$ 
takes us from $p(\w,\y|D,a,\tau^2)$ to $p(\w,\st|D,a,\tau^2)$, and setting $w_i=0$ when $s_i =-1$ undoes the replacement in (\ref{rjs}).

Since (\ref{ssn}) is a piecewise Gaussian distribution, we can sample from it again using the methods of~\cite{pakman2013exact}. 
As in that work, the possible truncations for $\w$ are given  as $g_n(\w) \geq 0$ for $n=1, \ldots, N$, with $g_n(\w)$ any product of linear and quadratic functions of $\w$.
The details are a simple extension of the purely binary case and are not very illuminating, so  we leave them for the Appendix.

\subsection{Probit regression}
Consider a probit regression model in which binary variables $b_i=\pm 1$ are observed with probability
\eqan
p(b_i|\w, \x_i) = \frac{1}{\sqrt{2\pi}}\int_{z_ib_i \geq 0} dz_i e^{-\frac12(z_i + \x_i\w )^2}
\enan
Given a set of $N$ pairs $(b_i, \x_i)$, we are interested in the posterior distribution of the weights $\w$ using the spike-and-slab 
prior~(\ref{sns}).
This posterior is the marginal over the $z_i$'s of the distribution
\eqan
p(\z,\w,\st|\x,a,\tau^2) \vp \prod_{i=1}^N e^{-\frac12(z_i + \x_i\w )^2} p(\w,\st|a,\tau^2) \quad \quad z_ib_i \geq 0 \,,
\enan
and we can use the same approach as above to transform this distribution into a truncated piecewise Gaussian, defined over 
the $(N+2d)$-dimensional space of the vector $(\z,\w,\y)$. Each $z_i$ is truncated according to the sign of $b_i$ and we can 
also  truncate the $\w$ space if we so desire. We omit the details of the HMC algorithm, since it is very similar to the linear regression 
case.

\section{Examples}
We present here three examples that illustrate the advantages of the proposed HMC algorithms over Metropolis or Gibbs samplers.
\subsection{1D Ising model}
\begin{figure}[t]
\begin{center}
\includegraphics[scale=0.53]{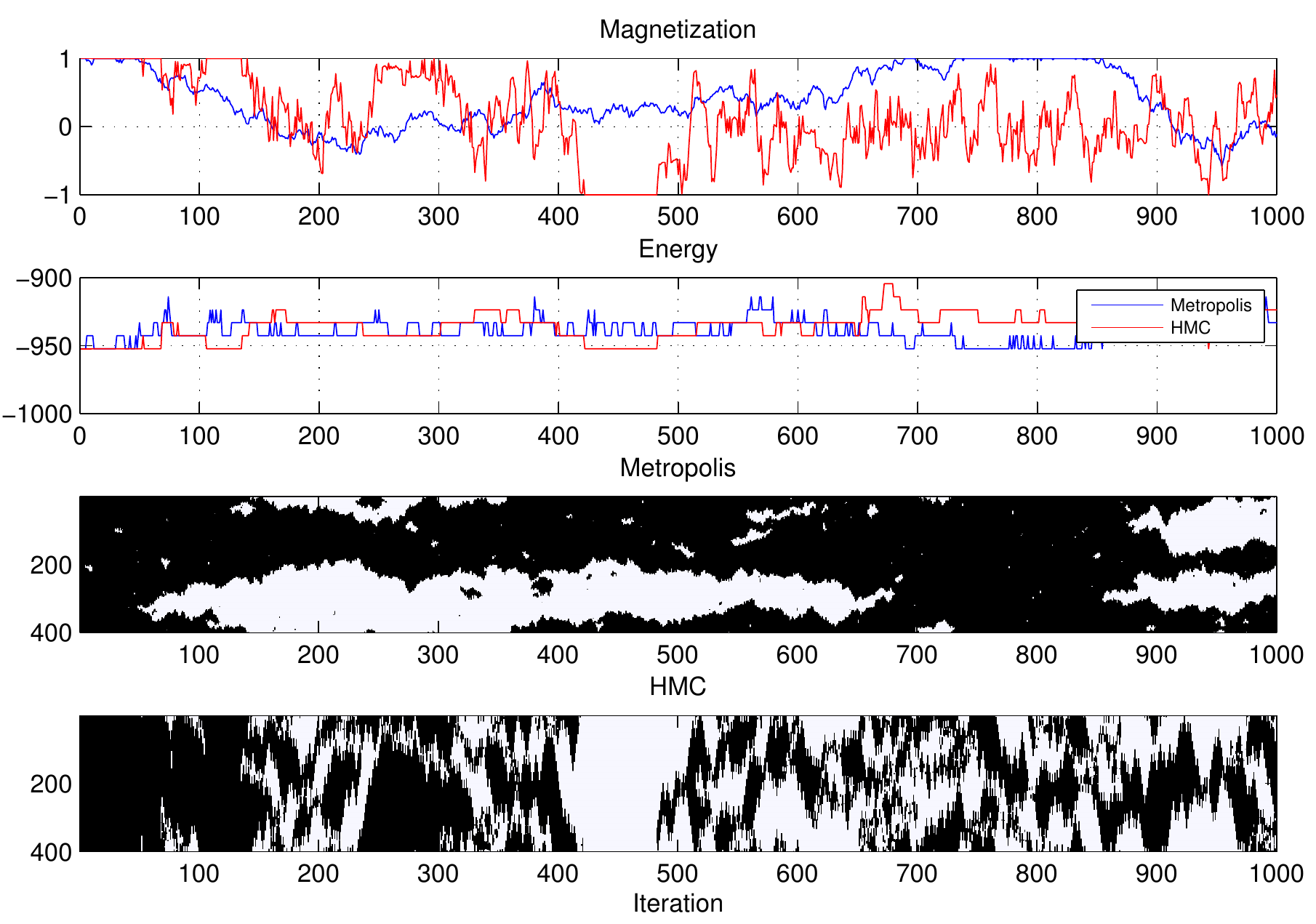}
\end{center}
\caption{ {\bf 1D Ising model.} 
First $1000$ iterations of Gaussian HMC and Metropolis samplers on a model with $d=400$ and $\beta=0.42$, initialized with all spins $s_i=1$ (black dots). 
For HMC the travel time was $T=12.5\pi$ and in the Metropolis sampler we recorded the state of the Markov chain once every $d \times 12.5$ flip proposals.
The lower two panels show the state of $\st$ at every iteration for each sampler. The plots show clearly that the HMC model mixes faster than Metropolis in this model.}
\label{ising1d}
\end{figure}

We consider a 1D periodic Ising model, with $p(\st) \vp e^{-\beta E[\st]}$, 
where the energy is $E[\st] = - \sum_{i=1}^d s_{i} s_{i+1}$, with $s_{d+1}=s_{1}$ and $\beta$ is the inverse temperature. Figure~\ref{ising1d} shows  the first $1000$ iterations of 
both the Gaussian HMC and the Metropolis\footnote{As is well known (see e.g.\cite{newman1999monte}), for binary distributions, the Metropolis sampler that 
chooses a random spin and makes a proposal of flipping its value, is more efficient than the Gibbs sampler. } 
sampler on a model with $d=400$ and $\beta=0.42$, initialized with all spins $s_i=1$. 
In HMC we took a travel time $T=12.5\pi$ and, for the sake of comparable 
computational costs, for the Metropolis sampler we recorded the value of $\st$  every $d \times 12.5$ flip proposals.
The plot shows clearly that the Markov chain mixes much faster with HMC than with Metropolis.
A useful variable that summarizes the behavior of the Markov chain is 
the magnetization 
$ m =  \frac{1}{d}\sum_{i=1}^d s_i  $\,,
whose expected value is $\langle m \rangle=0$. The oscillations of both samplers around this value illustrate the superiority of the HMC sampler.
In the Appendix we present a more detailed comparison of the HMC Gaussian and exponential  and the Metropolis samplers, showing that the Gaussian 
HMC sampler is the most efficient among the three.

\subsection{2D Ising model }
\begin{figure}[t]
\begin{center}
\includegraphics[scale=0.50]{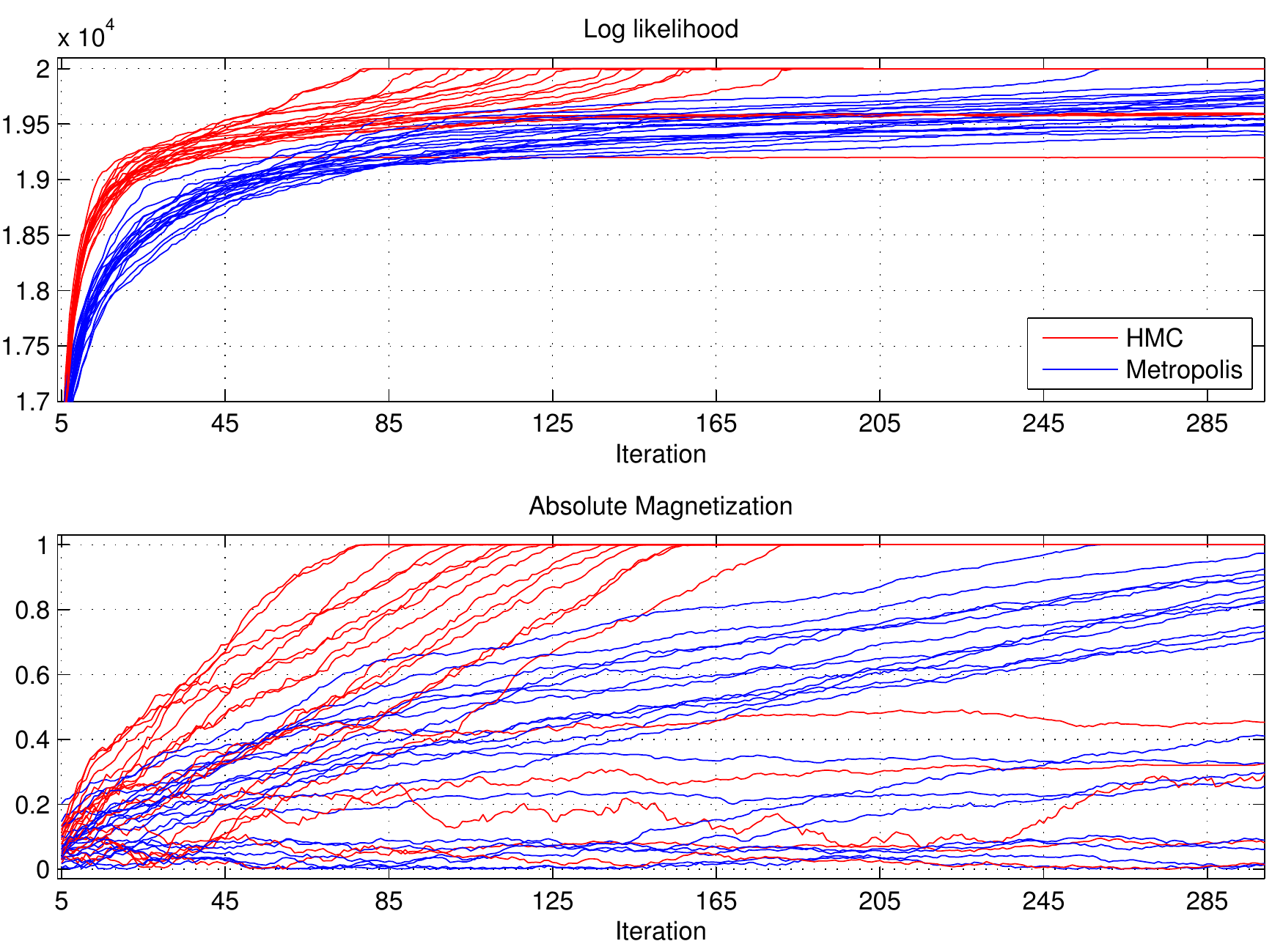}
\end{center}
\caption{ {\bf 2D Ising model.}  First samples from $20$ simulations in a 2D Ising model in a square lattice of side length $L=100$ with periodic boundary conditions
and inverse temperature~$\beta = 0.5$. The initial state is totally disordered. We do not show the first 4 samples in order to ease the visualization.
For the Gaussian HMC we used $T=2.5 \pi$ and for Metropolis we recorded 
the state of the chain  every $2.5 L^2 $ flip proposals. The plots illustrate that in general HMC reaches equilibrium faster than Metropolis in this model. }
\label{ising2d}
\end{figure}

We consider now a 2D Ising model on a square lattice of size $L\times L$ with periodic boundary conditions below the critical temperature. Starting from a completely disordered state, 
we compare the time it takes for the sampler to reach one of the two low energy states with magnetization $m \simeq \pm 1$.
Figure~\ref{ising2d} show the results of 20 simulations of such a model with $L=100$ and inverse temperature~$\beta=0.5$. 
We used a Gaussian HMC with $T=2.5 \pi$ and a Metropolis sampler recording values of $\st$ every~$ 2.5 L^2$ flip proposals.
In general we see that the HMC sampler reaches higher likelihood regions faster.

Note that these results of the 1D and 2D Ising models illustrate the advantage of the HMC 
method in relation to two different time constants relevant for Markov chains~\cite{sokal1989monte}. Figure~\ref{ising1d} shows 
that the HMC sampler explores faster the sampled space once the chain has reached its equilibrium distribution, while Figure~\ref{ising2d} shows that the HMC sampler
is faster in reaching the equilibrium distribution.

\subsection{Spike-and-slab linear regression with positive coefficients}
\begin{figure}[t]
\begin{center}
\includegraphics[scale=0.56]{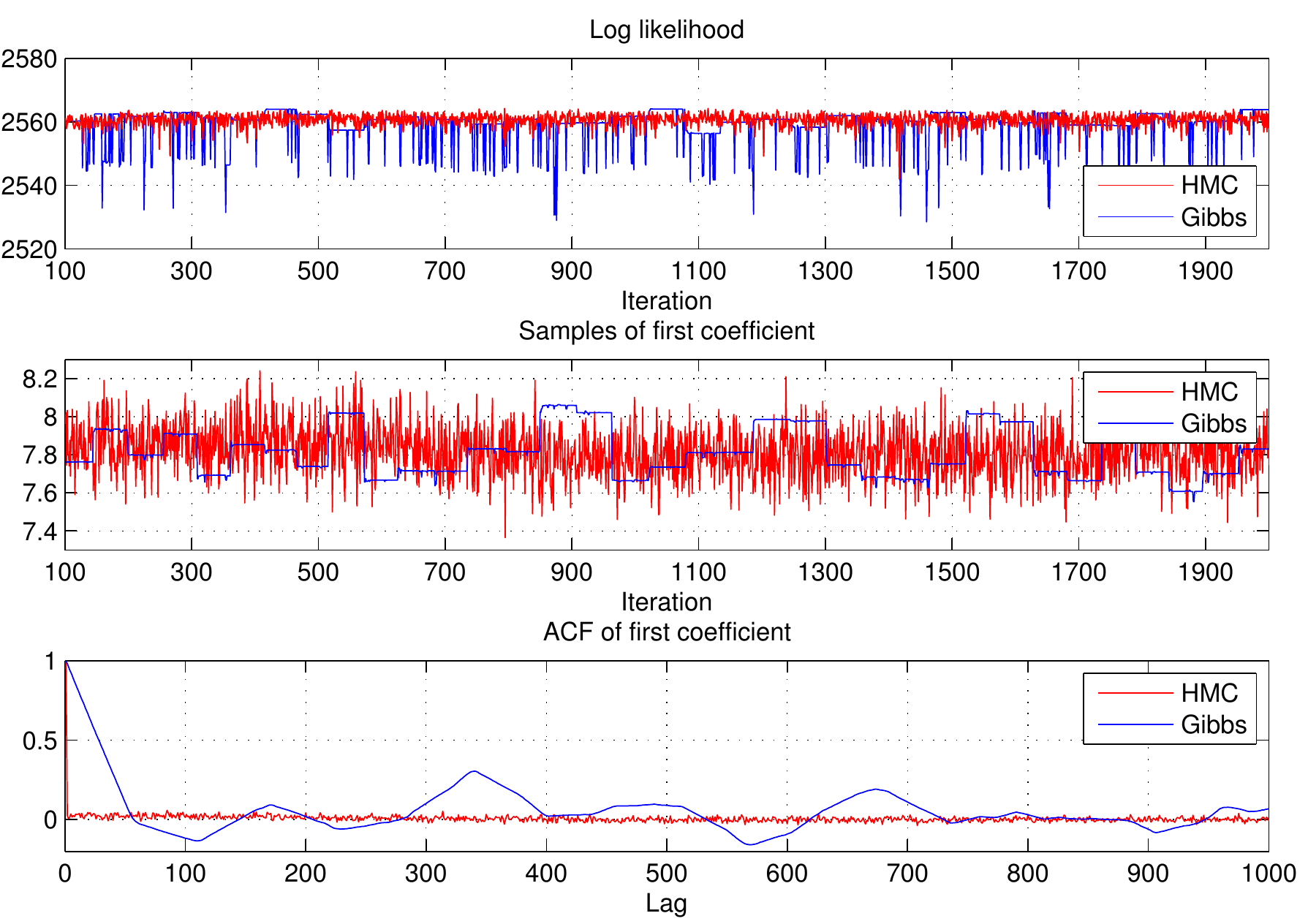}
\end{center}
\caption{ {\bf Spike-and-slab linear regression with constraints.}   
Comparison of the proposed HMC method with the Gibbs sampler of~\cite{mohamed2011bayesian} for the posterior of a linear regression model with spike-and-slab
prior, with a positivity constraint on the coefficients. See the text for details of the synthetic data used. 
Above: log-likelihood as a function of the iteration. Middle: samples of the first coefficient. Below: ACF of the first coefficient. 
The plots shows clearly that HMC mixes much faster than Gibbs and is more consistent in exploring areas of high probability.
}
\label{hmc_spike}
\end{figure}

We consider a linear regression model ${\bf z} = X\w + {\bf \varepsilon}$ with the following synthetic data. $X$ has  $N=700$ rows,
each sampled from a $d=150$-dimensional Gaussian whose covariance matrix has $3$ in the diagonal and $0.3$ in the nondiagonal entries. The noise is
${\bf \varepsilon} \sim {\cal N}(0,\sigma^2=100)$. The data ${\bf z}$ is generated with a coefficients vector $\w$, with 10 non-zero entries with values between $1$ and $10$.
The spike-and-slab hyperparameters are set to $a=0.1$ and $\tau=10$. Figure~\ref{hmc_spike} compares the results of the proposed HMC method versus the Gibbs
sampler used in~\cite{mohamed2011bayesian}. In both cases we impose a positivity constraint on the coefficients. 
For the HMC sampler we use a travel time $T=\pi/2$. This  results in a number of wall hits (both for $\w$ and $\y$ variables) of $\simeq 150$,
which makes the computational cost of every HMC and Gibbs sample similar.
The advantage of the HMC method is clear, both in exploring regions of higher probability and in the mixing speed  of the sampled coefficients.
This impressive difference in the efficiency of HMC versus Gibbs is similar to the case of truncated multivariate Gaussians studied 
in~\cite{pakman2013exact}.

\section{Conclusions and outlook}
We have presented a novel approach to use exact HMC methods to sample from generic binary distributions and certain distributions over mixed binary and continuous variables, 

Even though with the HMC algorithm is better than Metropolis or Gibbs in the examples we presented, 
this will clearly not be the case in many complex  binary distributions for which specialized sampling algorithms have been developed,
such as the Wolff or Swendsen-Wang algorithms for  2D Ising models near the critical temperature~\cite{newman1999monte}.
But in particularly difficult distributions, these HMC algorithms could be embedded as inner loops inside more powerful algorithms of Wang-Landau type~\cite{wang2001efficient}.
We leave the exploration of these newly-opened realms for future projects.


\subsubsection*{Acknowledgments}
This work was supported by an NSF CAREER award and by the US Army Research Laboratory and the US Army Research Office under contract number W911NF-12-1-0594.

{\bf \large Appendix}
\appendix
 \section{Wall-crossing rate in the Gaussian augmentation}
In the Gaussian augmentation, the equilibrium distribution of $(\y, \q)$ in each orthant is 
\eqan
p(\y,\q|\st) \vp e^{- \frac{\y \cdot \y}{2} } e^{- \frac{\q \cdot \q}{2} } \,,
\enan 
and therefore the distribution of 
\eqan 
u_i = y_i^2 + q_i^2    \qquad \qquad  i =1, \ldots, d.
\enan
is $\chi^2_2$, chi-squared with two degrees of freedom.  
Due to conservation of energy, each $u_i$ is constant while the particle stays in an orthant and only changes if it crosses the $y_i=0$ wall. 
When the particle hits the $y_i=0$ wall, we have $u_i=q_i^2(t_i^-)$, and the particle crosses  if
\eqan 
u_i > -2\log p (-s_i, \st_{-i}) + 2\log p (s_i, \st_{-i}) \,.
\enan 
The probability of this event is 
\eqan
P \left[ u_i > -2\log \left( \frac{p (-s_i, \st_{-i})} {p(s_i, \st_{-i})} \right) \right] = 
\left\{
 \begin{array}{ll}
1    &  \textrm{for} \,\,  \frac{p (-s_i, \st_{-i})} {p(s_i, \st_{-i})} > 1
\\
1-C_{\chi^2_2}(-2\log \left( \frac{p (-s_i, \st_{-i})} {p(s_i, \st_{-i})} \right))  &  \textrm{for} \,\,  \frac{p (-s_i, \st_{-i})} {p(s_i, \st_{-i})} < 1
\end{array}
\right.  
\label{pp}
\enan 
where 
\eqan
C_{\chi^2_2}(x) = 1-e^{-\frac{x}{2}}
\enan 
is the cdf of $\chi^2_2$. Inserting this expression in (\ref{pp}) gives
\eqan 
P \left[ u_i > -2\log \left( \frac{p (-s_i, \st_{-i})} {p(s_i, \st_{-i})} \right) \right] = \min\left(1, \frac{p (-s_i, \st_{-i})} {p(s_i, \st_{-i})} \right)
\enan 
which is exactly the probability of acceptance in  a Metropolis algorithm that 
samples uniformly a value for $i$ and makes a proposal of flipping the binary variable $s_i$.

\section{Comparing the efficiency of binary samplers}
\begin{figure}[!h]
\begin{center}
\includegraphics[scale=0.63]{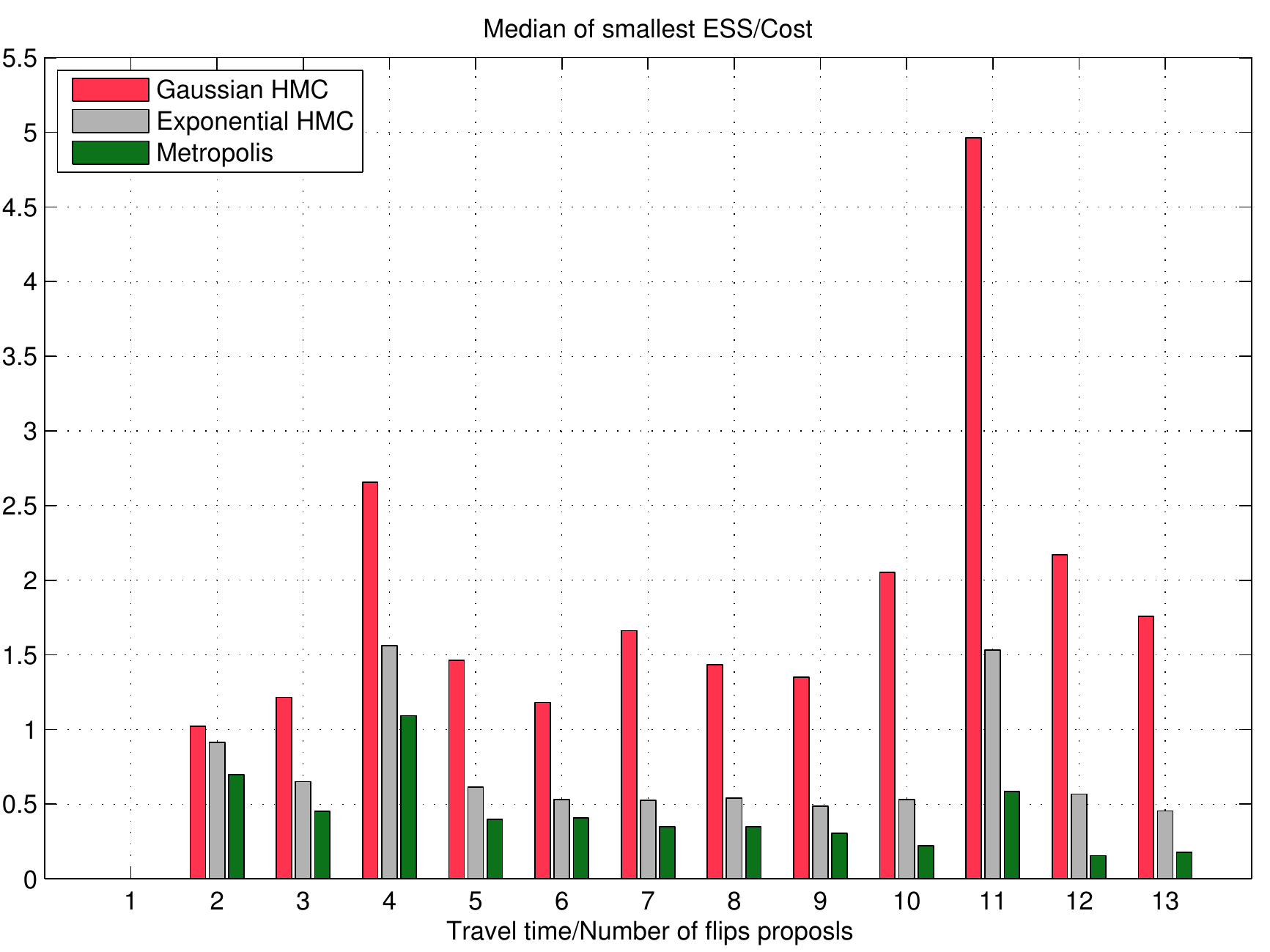}
\end{center}
\caption{ {\bf Efficiency comparison for binary samplers.}   
 We considered a 1D Ising model with $d=400$ and $\beta=0.42$. 
 In the Gaussian HMC sampler we considered $T=(n-1/2)\pi$ with $n=1,\ldots, 13$, for Metropolis we recorded the state of the chain after $d \times (n-1/2)$ flip proposals and for 
 the exponential HMC case we used $T$'s corresponding to similar computational costs. 
 For each $n$ and each sampler we took $3000$ samples and recorded the smallest effective sample size (ESS) among the  $s_i$ variables. 
 We repeated this 10 times and computed the median value of these smallest ESSs. The plot shows these values divided by the computational cost for each $n$.
Note that the HMC Gaussian sampler is consistently more efficient.
 }
\label{ess}
\end{figure}

We performed a more detailed comparison of the efficiency of the binary HMC sampler with Gaussian and exponential augmentations and the Metropolis sampler. 
As in Section $4.1$, we considered a 1D Ising model with $d=400$ and $\beta=0.42$. The results are in Figure~\ref{ess} and show that the 
HMC sampler with Gaussian  augmentation is the most efficient of the three samplers.

\section{Details of spike-and-slab linear regression with truncated parameters}  
We want to sample from the distribution
\eqan 
p(\w,\y|D,a,\tau^2)  \vp  e^{-\frac12  \w'_{+} \left( {\bf M}_{+} + \tau^{-2} \right) \w_{+}   + {\bf r}_{+}\cdot \w_{+} }            
           e^{-\frac{\w_{-}\cdot \w_{-} }{2 \tau^2}} e^{-\frac{\y \cdot \y}{2}}  a^{|\spl|} (1-a)^{|\sm|}
\enan 
where the values of $\st$ in the rhs are obtained from the signs of $\y$.
Since (\ref{ssn}) is a piecewise Gaussian distribution, we can sample from it using the methods of~\cite{pakman2013exact}. For this, we introduce
momentum variables $q_i$ and $g_i$ associated to the coordinates $y_i$ and $w_i$ and consider the Hamiltonian
\eqan 
H &=& H_{\y, \q} + H_{\w, \g} - |\spl|\log a - |\sm| \log (1-a)
\label{HH}
\\
H_{\y, \q} &=& \frac{\y \cdot \y }{2} + \frac{\q \cdot \q }{2} 
\\
H_{\w, \g} &=&  \frac{\w_{+}' {\boldsymbol \Sigma}_+^{-1} \w_{+}}{2} - {\bf r}_{+} \! \cdot \w_{+}  + \frac{\g_{+}' {\boldsymbol \Sigma}_+ \g_{+}}{2}
+ \frac{\w_{\!-} \!\!\cdot \w_{\!-}}{2 \tau^2} + \frac{\g_{-} \!\! \cdot \g_{-}}{2 \tau^{-2}} 
\\
\nn
&-& \frac12 \log|{\boldsymbol \Sigma}_+| - \frac12 |\sm| \log(\tau^2)
\enan 
where we defined 
\eqan 
{\boldsymbol \Sigma}_+ = \left({\bf M}_{+} + \tau^{-2}\right)^{-1}  \,.
\enan 
Note that we have chosen a mass  matrix for $\g$ that depends on the orthant of $\y$, much like the potential terms for $\w$.  
This choice leads to decoupled equations of motion for all the coordinates, with solutions
\eqan
y_i(t)  &=&  y_i(0) \cos(t) + q_i(0) \sin(t)  \,,
\label{yit}
\\
w_i(t) &=& \mu_i + (w_i(0) -\mu_i)\cos(t) + \dot{w}_i(0) \sin(t)   \,,
\label{wit}
\enan 
where in each orthant the components of ${\boldsymbol \mu}$ are
\eqan
{\boldsymbol \mu}_{-} &=& 0 \,,
\\
{ \boldsymbol \mu}_{+} &=& {\boldsymbol \Sigma}_+  {\bf r}_{+}     \,.
\label{mus}
\enan 
Each iteration of the sampling algorithm consists of sampling initial values for $\q$ and $\dot{\w}$ from 
\eqan
q_i(0) &\sim& {\cal N}(0,1) \,,
\\
\dot{w}_i(0) &\sim& {\cal N}(0,\tau^2) \qquad \textrm{for   } s_i = -1 \,,
\\
\dot{\w}_{+}(0)  &\sim& {\cal N}(0, {\boldsymbol \Sigma}_+) \,,
\enan 
and letting the particle move during a time $T$ according to the Hamiltonian (\ref{HH}). As before, the final coordinates belong 
to a Markov chain with invariant distribution $p(\w,\y|D,a\tau^2)$, and are used as the initial coordinates of the next iteration.
Note that it is more convenient to sample $\dot{\w}$ 
instead of $\g$ (related by $\dot{\w}_+= { \boldsymbol \Sigma}_{+} \g_{+}, \dot{\w}_-= \tau^2 \g_{-}$), because it is the former that appears in (\ref{wit}).

The trajectory of the particle in the $(\y,\w)$-space is given by (\ref{yit})-(\ref{wit}) until some coordinate $y_j$ reaches $y_j=0$ at time $t_j$, 
or, if the space of $\w$ is truncated, the $\w$ coordinates touch the boundary of their allowed space.  
Consider the first case and suppose that $y_j < 0$ for $t < t_j$. 
The conservation of energy across the $y_j=0$ boundary implies
\eqan 
\frac{q_j^2(t_j^+)}{2} = \Delta_j + \frac{q_j^2(t_j^-)}{2} \,,
\label{kin2}
\enan 
and the energy jump $\Delta_j$ depends  on  $\w$ and $\g$ and is given by 
\eqan
\Delta_j = - H_{\w, \g}(\st_{-j}, s_j=+1) + H_{\w, \g}(\st_{-j}, s_j=-1) + \log(a/(1-a)) \,.
\label{deltaj2}
\enan 
Note that the trajectory of $\w$, $\g$  is continuous at $t=t_j$, and (\ref{deltaj2}) only refers to the change 
in the functional form of $H$ across the boundary. If (\ref{kin2}) gives
a positive value for $q_j^2(t_j^+)$, the particle crosses the $y_j=0$ boundary, and if not, it bounces back with 
$q_j(t_j^+) = -q_j(t_j^-)$. 
In the $\w$-truncated case, when the  $\w$ coordinates touch the boundary of their allowed space, the velocity $\dot{\w}$ is reflected off 
the boundary in an elastic collision,  similarly to the truncated Gaussians discussed in~\cite{pakman2013exact}.




\bibliographystyle{unsrt}
\bibliography{hmc}

\begin{thebibliography}{10}

\bibitem{neal2011mcmc}
Radford Neal.
\newblock {MCMC Using Hamiltonian Dynamics}.
\newblock {\em Handbook of Markov Chain Monte Carlo}, pages 113--162, 2011.

\bibitem{pakman2013exact}
Ari Pakman and Liam Paninski.
\newblock {Exact Hamiltonian Monte Carlo for Truncated Multivariate Gaussians}.
\newblock {\em Journal of Computational and Graphical Statistics}, 2013,
  arXiv:1208.4118.

\bibitem{hertz1991introduction}
John~A Hertz, Anders~S Krogh, and Richard~G Palmer.
\newblock {\em Introduction to the theory of neural computation}, volume~1.
\newblock Westview press, 1991.

\bibitem{zhang2012continuous}
Yichuan Zhang, Charles Sutton, Amos Storkey, and Zoubin Ghahramani.
\newblock {Continuous Relaxations for Discrete Hamiltonian Monte Carlo}.
\newblock In {\em Advances in Neural Information Processing Systems 25}, pages
  3203--3211, 2012.

\bibitem{hoffman2011no}
M.D. Hoffman and A.~Gelman.
\newblock {The No-U-Turn sampler: adaptively setting path lengths in
  Hamiltonian Monte Carlo}.
\newblock {\em Arxiv preprint arXiv:1111.4246}, 2011.

\bibitem{park2008bayesian}
T.~Park and G.~Casella.
\newblock {The Bayesian lasso}.
\newblock {\em Journal of the American Statistical Association},
  103(482):681--686, 2008.

\bibitem{carvalho2010horseshoe}
C.M. Carvalho, N.G. Polson, and J.G. Scott.
\newblock The horseshoe estimator for sparse signals.
\newblock {\em Biometrika}, 97(2):465--480, 2010.

\bibitem{mitchell1988bayesian}
T.J. Mitchell and J.J. Beauchamp.
\newblock Bayesian variable selection in linear regression.
\newblock {\em Journal of the American Statistical Association},
  83(404):1023--1032, 1988.

\bibitem{george1993variable}
E.I. George and R.E. McCulloch.
\newblock {Variable selection via Gibbs sampling}.
\newblock {\em Journal of the American Statistical Association},
  88(423):881--889, 1993.

\bibitem{mohamed2011bayesian}
S.~Mohamed, K.~Heller, and Z.~Ghahramani.
\newblock {Bayesian and L1 approaches to sparse unsupervised learning}.
\newblock {\em arXiv:1106.1157, International Conference on Machine Learning
  (ICML)}, 2012.

\bibitem{goodfellow2012spike}
I.J. Goodfellow, A.~Courville, and Y.~Bengio.
\newblock Spike-and-slab sparse coding for unsupervised feature discovery.
\newblock {\em arXiv:1201.3382, International Conference on Machine Learning
  (ICML)}, 2012.

\bibitem{chen2012bayesian}
Yutian Chen and Max Welling.
\newblock {Bayesian structure learning for Markov random fields with a spike
  and slab prior}.
\newblock {\em arXiv:1206.1088, Conference on Uncertainty in Artificial
  Intelligence}, 2012.

\bibitem{green1995reversible}
Peter~J Green.
\newblock {Reversible jump Markov chain Monte Carlo computation and Bayesian
  model determination}.
\newblock {\em Biometrika}, 82(4):711--732, 1995.

\bibitem{newman1999monte}
Mark~E.J. Newman and Gerard~T. Barkema.
\newblock {Monte Carlo methods in statistical physics}.
\newblock {\em { Oxford: Clarendon Press, 1999.}}, 1, 1999.

\bibitem{sokal1989monte}
Alan~D Sokal.
\newblock { Monte Carlo methods in statistical mechanics: foundations and new
  algorithms}, 1989.

\bibitem{wang2001efficient}
Fugao Wang and David~P Landau.
\newblock Efficient, multiple-range random walk algorithm to calculate the
  density of states.
\newblock {\em Physical Review Letters}, 86(10):2050--2053, 2001.

\end{thebibliography}

\end{document}